# Comment on "Bilayer aggregate microstructure determines viscoelasticity of lung surfactant suspensions" by C.O. Ciutara and J.A. Zasadzinski, Soft Matter, 2021, 17, 5170-5182

Jean-François Berret*

Université Paris Cité, CNRS, Matière et Systèmes Complexes, 75013 Paris, France

**Abstract** : For applications of pulmonary surfactant delivery to the lungs, the question of rheology of the existing clinical formulations is of upmost importance. Recently, Ciutara and Zasadsinky (Soft Matter, 2021, 17, 5170-5182) measured the rheological properties of Infasurf®, Survanta® and Curosurf®, three of the most used pulmonary surfactant substitutes. This study revealed that these fluids are shear-thinning and characterized by a yield stress. The results obtained by Ciutara *et al.* on Curosurf® differ from our results published in Colloids and Surfaces B: Biointerfaces, 2019, 178, 337-345 and in ACS Nano, 2020, 14, 466 - 475. In contrast, we found that Curosurf® suspensions are viscous Newtonian or slightly shear-thinning fluids, with no evidence of yield stress. The purpose of this Comment is to discuss possible causes for the discrepancy between the two studies, and to suggest that for biological fluids such as surfactant substitutes, the microrheology technique of rotational magnetic spectroscopy (MRS) can provide valuable results.

*Corresponding authors: jean-francois.berret@u-paris.fr
Website: https://www.jean-francois-berret-website-pro.fr/

*Introduction*: In the lungs, the alveoli form a froth-like structure and provide a large area that favors the transfer of the oxygen into the bloodstream and the carbon dioxide from the bloodstream into the exhaled air.[1,2] In mammals, the alveoli are constituted by an epithelium wall of thickness around 1 µm. One side of this epithelium is covered with lung lining fluid, also known as pulmonary surfactant and on the other side it is surrounded by blood capillaries. The pulmonary surfactant is made of phospholipids and proteins in a ratio 90:10 at the physiological concentration of 40 g L$^{-1}$.[3-6] The role of surfactant is to reduce the surface tension with the air and to prevent alveolar collapse and expansion during breathing [7,8] With regard to the inhalation of airborne nanoparticles, the pulmonary surfactant also represents an active protective barrier against external particles and pathogens.[9,10] In practice only few studies have investigated human or animal native (i.e. endogenous) pulmonary fluids, the reason being that they are difficult to extract, purify and condition to obtain significant reproducible data.[2,8] Researchers have turned to pulmonary surfactant substitutes (i.e. exogenous), such as Infasurf®, Survanta® and Curosurf®, which are medications for treatment of preterm infants with acute respiratory distress syndrome (ARDS), a syndrome associated with the hyaline membrane disease.[11,12] The structural properties of exogenous surfactant are now well documented. Curosurf® for instance has been studied in details, and electron and optical microscopy have shown that the lipids associate into membranes to form multi-lamellar and multivesicular vesicles with size between 50 nm to 5 µm.[13-15] In contrast, their flow properties are much less known. However, these properties are critical in surfactant replacement strategies for children and adults[11,12,16] where fluid velocity in successive bifurcations can vary from 1 m s$^{-1}$ in the trachea to 10$^{-3}$ m s$^{-1}$ in the distal region of the lungs.[17] Given the importance of the rheological properties of lung fluids for biomedical applications, it is surprising that only few rheology experiments have been performed yet.[18-20]

Recently, Ciutara *et al.* measured the flow properties of Infasurf®, Survanta® and Curosurf® using cone-and-plate rheometry.[21] These authors were able to produce a consistent set of data showing that the three fluids were shear-thinning and characterized by a yield stress ($\sigma_Y$), implying that these fluids flow if a stress applied to the sample is greater than $\sigma_Y$.[22-24] In $\eta(\sigma)$-flow curves, where $\eta(\sigma)$ is the shear viscosity at the applied shear stress $\sigma$, the existence of a threshold stress is seen by a sharp decrease in viscosity at $\sigma = \sigma_y$ (Figure 2 from Ref.[21]). The yield stress values obtained were on the order of $\sigma_y$ = 0.02 Pa for Infasurf® and Curosurf®, and $\sigma_y$ = 1 Pa for Survanta®. The existence of residual elasticity at low frequencies has also been confirmed by the authors in linear rheology measurements, where the elastic and loss moduli $G'(\omega)$ and $G''(\omega)$ show scaling laws as a function of the frequency with exponents around 0.4, together with the inequality $G'(\omega) > G''(\omega)$.

The results obtained by Ciutara *et al.* on Curosurf® differ from our results acquired by cone-and-plate rheometry and by microrheology techniques.[25] The microrheology experiments were performed at Curosurf® concentrations between 5 and 80 g L$^{-1}$ using remotely operated magnetic wires submitted to a rotating field. As described below, monitoring the wire behavior as a function of frequency provides access to the static shear viscosity and elastic modulus. These results were published in 2019 in the reference:[25] L.P.A. Thai et al., *Colloids and Surfaces B: Biointerfaces* **178**, 337-345 (2019). In this report, we find, in contrast to Ciutara *et al.*[21] that Curosurf® suspensions are viscous Newtonian or slightly shear-thinning fluids (depending on the concentration), with no evidence of a yield stress behavior. For applications of mimetic surfactant delivery, the question of rheology and in particular whether or not they are yield stress fluids is paramount. Such applications include surfactant replacement strategies applied to adults suffering from COVID-19 infection[16,26,27] or less invasive surfactant administration (LISA) technologies for preterm infants.[28] In what follows, we first compare our macroscopic rheological data with those of Ciutara *et al.*[21] and recall the salient features of microrheology before concluding on the rheology of Curosurf®.

***Comparison of rheological data from Ciutara et al.[21] and Thai et al.[25]***:
Fig. 1 compares the results obtained by the two groups in dynamical frequency sweeps for $G'(\omega)$ and $G''(\omega)$ and at 40 and 80 g L$^{-1}$. All $\omega$-dependence can be accounted for by scaling laws of the form $G'(\omega) \sim \omega^\alpha$ and $G''(\omega) \sim \omega^\beta$, as indicated by the straight lines fitting the data points. In Fig. 1, the exponents $\alpha$ and $\beta$ equal 0.39 and 0.43 for the 80 g L$^{-1}$ data from Ciutara *et al.*,[21] and $\beta$ equals 0.95 and 0.75 from Thai *et al.* for 40 and 80 g L$^{-1}$, respectively.[25] Regarding $G''(\omega)$, the power law exponents and magnitudes differ, except at the highest frequency (~ 10 rad s$^{-1}$) where they appear to intercept. The storage modulus $G'(\omega)$ was also measured in the work of Thai *et al.*[25] In **Supplementary Information S1**, we show the strain (**Fig. S1a**) and frequency (**Fig. S1b**) sweep results obtained on Curosurf® 80 g L$^{-1}$. In **Fig. S1b**, $G'(\omega)$ is found to vary between 2×10$^{-3}$ and 2×10$^{-2}$ Pa in the range 0.1-10 rad s-1 and it is lower than $G''(\omega)$. For G'(ω), data above 10 rad s-1 were removed from the plot, because in this range, G'(ω) was found to decrease with frequency. Because the $G'(\omega)$-data was not available across the entire frequency domain, it was decided not to include these data in the Thai *et al*. 2019 paper[25] or in the main text of this Comment. Taken together, both strain and frequency sweeps, in the ranges 0.5-100% and 0.1-10 rad s$^{-1}$ show that Curosurf® 80 g L$^{-1}$ is viscoelastic and characterized by a linear response dominated by the viscous part of the complex elastic modulus, i.e. $G''(\omega) > G'(\omega)$.

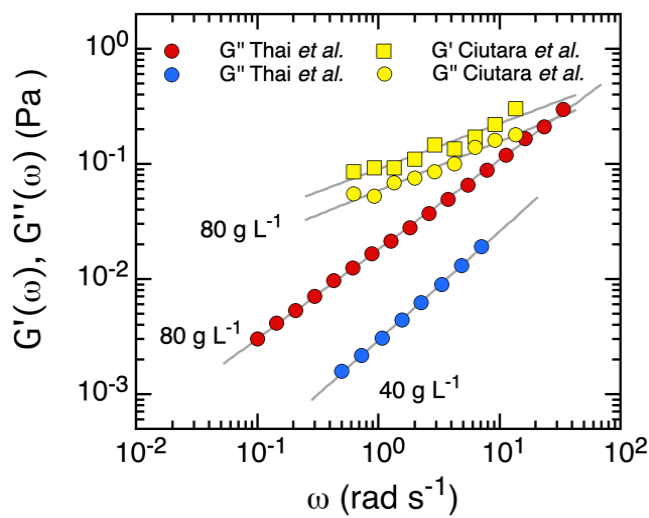

**Figure 1** : *Compilation of data of $G'(\omega)$ and $G''(\omega)$ measured by Thai et al. (T = 25 °C) [25] and by Ciutara et al. (T = 37 °C)[21] on Curosurf® suspensions at 40 and 80 g L$^{-1}$.The data in yellow are those of Fig. 5 in Ref.[21]*

An additional discrepancy between the two papers concerns steady shear data, where stress and viscosity are measured as a function of shear rate $\dot{\gamma}$. Figs. 2a and 2c displays the stress *versus* shear rate $\sigma(\dot{\gamma})$ curves obtained for Curosurf® 80 g L$^{-1}$ at 25 °C and 37 °respectively, whereas Figs. 2b and 2d show the corresponding viscosity $\eta(\dot{\gamma})$-data. In Ciutara *et al.*,[21] steady shear rheology was performed as a function of $\dot{\gamma}$ over the range 10$^{-2}$ – 10$^{3}$ s$^{-1}$, the waiting time between each subsequent shear rate being 10 seconds. In Thai *et al.*,[25] the stress *versus* shear rate curve was measured following first a ramp of velocity gradients starting from 10$^{-2}$ to 1000 s$^{-1}$ (red dots), the entire scan lasting 15 minutes (measuring time 10 seconds and waiting time 20 seconds). At 1000 s$^{-1}$, the shear rate was decreased down to 10$^{-2}$ s$^{-1}$ under similar waiting and measurement time conditions (blue dots). Following the approach suggested by Ovarlez and coworkers for complex fluids,[22,29] only the downward ramp data were considered in the Thai *et al.* paper,[25] as they allow overcoming the transient responses between two consecutive rates more easily than using an upward ramp. Moreover, in the published data (Fig. 2b in Ref.[25]), only viscosity and stress values that were identical on the upward and downward ramps were retained, limiting the data to the 1 - 1000 s$^{-1}$ region. An interesting result here is that there is a reasonable overlap of the $\sigma(\dot{\gamma})$ and $\eta(\dot{\gamma})$-data between the measurements by Thai *et al.*[25] and Ciutara *et al.*,[21] typically above 5 s$^{-1}$. The differences between the upward and downward ramps suggest the existence of transient responses in the low shear rate domain, reminiscent of results found in the surfactant phases of wormlike micelles[30] and bilayers.[31-33] Note that although Curosurf® membrane has the gel-to-fluid transition temperature at 29.5 °C (**Supplementary Information S2**), the Curosurf® viscosities at 25 and 37°C exhibit similar shear-thinning behaviors above 1 s$^{-1}$, indicating a weak temperature dependence (**Supplementary Information S3**). In conclusion of this part, we cannot exclude that the differences between the measurements in the two groups are due to the different protocols used. The results of both studies are close to each other for $G''(\omega)$ at high angular frequency and for the shear stress at high shear rates, but differ in the low frequency and shear rate regions.

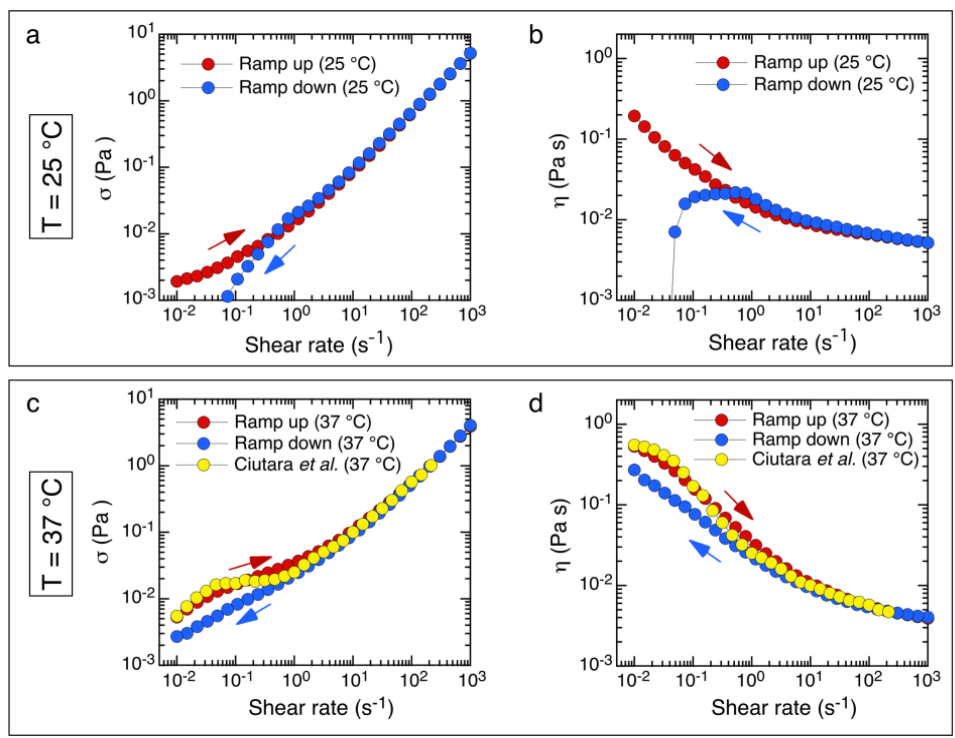

**Figure 2 : a, b)** *Shear stress and viscosity versus shear rate curves for Curosurf® 80 g L$^{-1}$ at T = 25 °C. **c,d)** Same as in a) and b) for T = 37 °C. The data in red (resp. blue) are from Thai et al.[25] following a ramp of velocity gradients from 10$^{-2}$ to 1000 s$^{-1}$ (resp. from 1000 s$^{-1}$ to 10$^{-2}$ s$^{-1}$), whereas the data in yellow are from Ciutara et al.[21]*

**Magnetic wire active microrheology data on Curosurf® as a function of the lipid concentration**
Recently, we applied the magnetic rotational spectroscopy (MRS) technique to determine the viscoelastic properties of complex fluids such as surfactant wormlike micelles,[34] polysaccharide gels,[35] the cytoplasm of living cells,[36] DNA strands complexed with amyloid fibers[37] and pulmonary mucus.[38] As mentioned above, it has also been applied to Curosurf® samples as a function of the concentration and in the presence of engineered nanoparticles, as shown in Thai et al, *ACS Nano* **14**, 466 - 475 (2020).[39] With MRS, magnetic wires of length between 1 and 100 µm, and of diameters 200 nm to 2 µm are submitted to a rotational magnetic field at increasing angular frequency $\omega$ and the wire motion is monitored by time-lapse microscopy. In its current configuration, the MRS allows to measure viscosity between 10$^{-3}$ and 10$^3$ Pa s and elasticity from 0.1 to 100 Pa over a frequency range between 10$^{-3}$ and 100 rad s$^{-1}$. The generic model of magnetic wire embedded in a viscous Newton liquid of static shear viscosity (**Fig. 3a**) $\eta$ predicts that at low frequency the wire rotates in phase with the field, leading to a synchronous motion.[40] With increasing $\omega$, the friction torque increases and above the critical value $\omega_C$ the wire undergoes a transition towards an asynchronous regime at:[40]

$$\omega_C = \frac{3}{8\mu_0} \frac{\Delta\chi}{\eta} \frac{B^2}{L^{*2}} \qquad (1)$$

In Eq. 1, $\mu_0$ is the vacuum permeability, $\Delta\chi$ the anisotropy of susceptibility between parallel and perpendicular wire direction, $B$ the magnetic field and $L^* = L/D\sqrt{g(L/D)}$, with $g(x) = ln(x) - 0.662 + 0.917/x - 0.050/x^2$.[34] With MRS, the simplest method to retrieve $\eta$ is to measure the critical frequency $\omega_C$ as a function of the reduced wire length $L^*$ and to verify the $1/L^{*2}$-dependence. The static viscosity is then determined from the prefactor $3\Delta\chi B^2/8\mu_0\eta$.

We further developed the concept of rotating wires applying it in soft solids such as crosslinked polysaccharide gels.[35] For soft solids, the critical frequency $\omega_C$ is identically zero whatever the frequency, and the wires oscillates between two symmetric orientations of amplitudes $\theta_B(\omega)$. This behavior can be understood with the help of Eq. 1 by assuming an infinite value for the viscosity, or equivalently by the fact that the stress applied is lower than that of the yield stress, the material being solely subjected to a deformation. In the asymptotic low and high frequency regimes, one gets for soft solids:[35]

$$\lim_{\omega \to 0} \theta_B(\omega) = \theta_{eq} \text{ and } \lim_{\omega \to \infty} \theta_B(\omega) = \frac{\theta_0 \theta_{eq}}{\theta_0 + \theta_{eq}} \qquad (2a)$$

$$\text{where } \theta_{eq} = 3\Delta\chi B^2/4\mu_0 G_{eq} L^{*2} \text{ and } \theta_0 = 3\Delta\chi B^2/4\mu_0 G L^{*2} \qquad (2b)$$

where $G_{eq}$ and $G$ denote the equilibrium and static storage elastic moduli, respectively. Table 1 summarizes the generic behaviors of a magnetic wire subjected to a rotating field as a function of frequency when placed in one of the four basic models of rheology, i.e. those of Newton, Maxwell, the Standard Linear Solid and Hooke.[41,42] The signatures of each model are clearly established, and show that the MRS method can differentiate between a viscous fluid and a soft solid. This can be achieved by studying the occurrence of the synchronous/asynchronous transition (column 4, Table 1) or the oscillation amplitude as a function of frequency (column 5, Table 1). We will use the predictions in Table 1 to identify the rheological nature of the Curosurf® (**Fig. 3b**).

**Fig. 3c** displays the ratio $8\mu_0\eta\omega_C/3\Delta\chi B^2$ as a function of the reduced wire length $L^*$ for Curosurf® samples at concentration 5, 20, 40, 50, 70 and 80 g L$^{-1}$. The data points are found to collapse on a single master curve displaying the $1/L^{*2}$-dependence (straight line). This set of results shows that according to Table I, all the samples studied obey a Newton or Maxwell type behavior. The data on the oscillation amplitude in Ref.[25] confirm that at physiological concentration $\theta_B(\omega) \sim \omega^{-1}$ and that $\lim_{\omega \to \infty} \theta_B(\omega) = 0$, suggesting finally a Newton-like behavior. At the concentration of the clinical formulation (80 g L$^{-1}$), $\theta_B(\omega)$ was found to deviate slightly from the Newton prediction, an outcome that was interpreted as the onset of viscoelasticity and could also be linked to the shear-thinning behavior in steady shear.

| Rheological model | Static shear viscosity | Yield stress | Sync./Async. transition | $\omega_C$ | $\lim_{\omega \to \infty} \theta_B(\omega)$ |
|---|---|---|---|---|---|
| **Newton** | $\eta$ | No | Yes | Eq. 1 | 0 |
| **Maxwell** | $\eta = G\tau$ | No | Yes | Eq.1 | $\theta_0$ |
| **Standard Linear Solid** | infinite | Yes | No | 0 | $\theta_0 \theta_{eq} / (\theta_0 + \theta_{eq})$ |
| **Hooke** | infinite | Yes | No | 0 | $\theta_{eq}$ |

*Table 1*: Summary of the predictions for the wire rotation in viscous (Newton) and viscoelastic (Maxwell) liquids, and in viscoelastic (Standard Linear Solid) and elastic (Hooke) solids.[41,42]

The synchronous-asynchronous transition observed ion Curosurf® 40 g L$^{-1}$ can be viewed in the Supporting Information movies in Refs.[25,39]. The values of the static viscosity used to construct the master diagram in **Fig. 3c** are displayed in **Fig. 3d** as a function of the Curosurf® concentration. At low concentration, $\eta(c)$ is increases linearly (dotted line) and deviates from the linear behavior above 40 g L$^{-1}$. The continuous line through the data points was obtained using a modified Krieger-Dougherty equation,[43] $\eta(c) = \eta_S(1 - c/c_m)^{-2}$ where $\eta_S$ is the solvent viscosity, $c_m$ the concentration at which the viscosity diverges and where the exponent in the function $(1 - c/c_m)$ was set to -2.[44] For Curosurf®, we found $c_m$ = 117 g L$^{-1}$. Assuming that $c_m$ is related to the maximum-packing volume fraction $\phi_m$ of vesicles and that its value for dispersed colloids is 0.63,[45,46] we end up with a volume fraction for the Curosurf® formulated at 80 g L$^{-1}$ of $\phi$ = 0.42, in good agreement with recent literature determinations.[20,21] Fitting the viscosity data using the Krieger-Dougherty three-parameter model[43] yields a similar continuous line in **Fig. 3d**, with $c_m$ = 101 g L$^{-1}$ and an exponent equal to 1.54. As a conclusion to this section, we have found that the results of microrheology on Curosurf®, whether as a function of frequency or concentration, are consistent with a description in terms of colloidal suspensions that is below the critical volume fraction $\phi_m$. In contrast to the paper by Ciutara *et al.*,[21] all studied suspensions were of Newton type (or slightly viscoelastic)) and no yield stress behavior could be evidenced.

In the course of this work, we questioned the ability of the MRS technique to detect low yield stresses (< 1 Pa), such as those reported by Ciutara *et al.*[21] for mimetic surfactants. In practice, we cannot exclude cases where the wires driven by the rotating magnetic field would disrupt the vesicle network, and locally induce flow. To answer this question, we return to the recent study we did on the effects of engineered nanoparticles on Curosurf® rheology. In this work, we studied the effects of 40 nm cationic silica (SiO$_2$) and alumina (Al$_2$O$_3$) nanoparticles on Curosurf® properties formulated in physiological conditions. It was found that particles interact strongly with Curosurf® vesicles, inducing profound modifications in the surfactant flow properties. In particular, we could show that above the alumina concentration of 0.1 g L$^{-1}$, Curosurf® underwent a rheological transition, from a purely viscous liquid to a soft solid behavior (described by the Standard Linear Model in Table 1). The transition was interpreted in terms of a bond percolation network, the particles acting as cross-links to the vesicles. Interestingly, the equilibrium elastic modulus $G_{eq}$ in Al$_2$O$_3$-Curosurf® mixtures (Eq. 2) was found in the range 0.01 - 0.1 Pa, and increasing with the particle concentration. For Curosurf® at 80 g L$^{-1}$, Ciutara *et al.* found low frequency storage modulus of the order of 0.1 Pa (Fig. 1), suggesting that if the Curosurf® formulation studied in Refs. [25,39] had been a soft solid with a finite yield stress, the MRS technique should have be able to detect it, which was not the case. These latter results represent a further indication of the differences in rheological response of Curosurf® studied by Thai. *et al.*[25] and by Ciutara *et al.*[21].

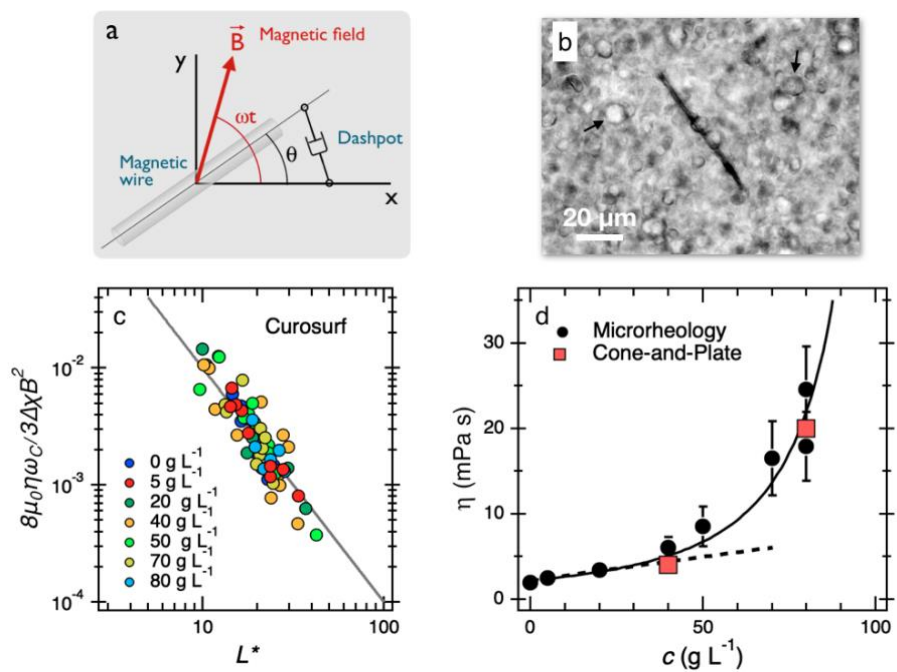

***Figure 3***: ***a)*** *Schematic representation of a wire actuated in a Newton viscous liquid represented by a dashpot. In MRS, the magnetic field rotates at a constant angular frequency ω.* ***b)*** *Phase-contrast optical microscopy image of a 40 g $L^{-1}$ Curosurf® dispersion at 25 °C (objective 20×) showing a 64 μm long magnetic wire. Arrows are pointing to vesicles.* ***c)*** *Normalized critical frequency $8\mu_0\eta\omega_C/3\Delta\chi B^2$ as a function of the reduced wire length $L^*$ obtained for Curosurf® suspensions at concentration 0 to 80 g $L^{-1}$ (T = 25 °C).[25] The straight line displays the $1/L^{*2}$-dependence predicted for Newton fluids (Eq.1).* ***d)*** *Concentration dependence of the Curosurf® static viscosity. The dotted line is from the Einstein model while the continuous line is derived from a modified Krieger-Dougherty equation.[44]*

***Conclusion***: We have shown in this Comment that from the macroscopic rheology and active microrheology data operating on the MRS principle, the Curosurf® samples showed, at all concentrations studied, a Newtonian or slightly viscoelastic behavior, in contradiction with the results from Ciutara *et al.*[21] who proposed that pulmonary surfactants are yield stress fluids. The concentrated Curosurf® suspension shows flow curves that suggest transient responses at low shear rates, as illustrated by the differences between the up- and downward ramps in **Fig. 2**. It should be recalled that many complex self-assembled fluids, including surfactant lamellar phases show transient behavior at low velocity gradients that can last for from minutes to hours.[31-33] Step shear rate experiments over extended periods of time are well suited to monitor these transient responses. They are however difficult to perform on Curosurf® because of the large amount of sample required. Here we show that difficulties inherent to classical rheometry, and in particular with regard to yield stress identification,[22-24] can be solved using our wire-based microrheology technique.

# Acknowledgments

We thank Mostafa Mokhtari and his team from the neonatal service at Hospital Kremlin-Bicêtre, Val-de- Marne, France for providing us the Curosurf® suspensions and Sandra Lerouge, Jesus Perez-Gil and Christophe Lenclud for fruitful discussions. Imane Boucema is acknowledged for letting us use the Anton Paar rheometer for the cone-and-plate rheology. This research was supported in part by the Agence Nationale de la Recherche under the contract ANR-12-CHEX-0011 (PULMONANO) and ANR-17-CE09-0017 (AlveolusMimics).

# References

1. P. Bajaj, J. F. Harris, J. H. Huang, P. Nath and R. Iyer, *ACS Biomater. Sci. Eng*, 2016, **2**, 473-488.
2. R. H. Notter, *Lung Surfactant: Basic Science and Clinical Applications*, CRC Press, Boca Raton, FL, 2000.
3. N. Hobi, G. Siber, V. Bouzas, A. Ravasio, J. Perez-Gil and T. Haller, *Biochim. Biophys. Acta, Biomembr.*, 2014, **1838**, 1842-1850.
4. T. Kobayashi, A. Shido, K. Nitta, S. Inui, M. Ganzuka and B. Robertson, *Resp. Physiol.*, 1990, **80**, 181-192.
5. J. F. Lewis and A. H. Jobe, *Am. Rev. Respir. Dis.*, 1993, **147**, 218-233.
6. M. Numata, P. Kandasamy and D. R. Voelker, *Expert Rev. Respir. Med.*, 2012, **6**, 243-246.
7. C. Casals and O. Canadas, *Biochim. Biophys. Acta, Biomembr.*, 2012, **1818**, 2550-2562.
8. E. Lopez-Rodriguez and J. Perez-Gil, *Biochim. Biophys. Acta, Biomembr.*, 2014, **1838**, 1568-1585.
9. A. Hidalgo, A. Cruz and J. Perez-Gil, *Biochim. Biophys. Acta, Biomembr.*, 2017, **1859**, 1740-1748.
10. M. Radiom, M. Sarkis, O. Brookes, E. K. Oikonomou, A. Baeza-Squiban and J. F. Berret, *Sci. Rep.*, 2020, **10**.
11. T. Curstedt, H. L. Halliday and C. P. Speer, *Neonatology*, 2015, **107**, 321-329.
12. J. Egberts, J. P. de Winter, G. Sedin, M. J. de Kleine, U. Broberger, F. van Bel, T. Curstedt and B. Robertson, *Pediatrics*, 1993, **92**, 768-774.
13. D. Waisman, D. Danino, Z. Weintraub, J. Schmidt and Y. Talmon, *Clin. Physiol. Funct. Imaging*, 2007, **27**, 375-380.
14. C. Schleh, C. Muhlfeld, K. Pulskamp, A. Schmiedl, M. Nassimi, H. D. Lauenstein, A. Braun, N. Krug, V. J. Erpenbeck and J. M. Hohlfeld, *Respir. Res.*, 2009, **10**, 90.
15. F. Mousseau, C. Puisney, S. Mornet, R. Le Borgne, A. Vacher, M. Airiau, A. Baeza-Squiban and J. F. Berret, *Nanoscale*, 2017, **9**, 14967-14978.
16. C. Lenclud, Curosurf® in Adult Acute Respiratory Distress Syndrome Due to COVID-19 (Caards-1), https://clinicaltrials.gov/ct2/show/NCT04384731.
17. A. Kazemi, B. Louis, D. Isabey, G. F. Nieman, L. A. Gatto, J. Satalin, S. Baker, J. B. Grotberg and M. Filoche, *PLoS Comput. Biol.*, 2019, **15**, e1007408.
18. D. M. King, Z. D. Wang, J. W. Kendig, H. J. Palmer, B. A. Holm and R. H. Notter, *Chem. Phys. Lipids*, 2001, **112**, 11-19.
19. D. M. King, Z. D. Wang, H. J. Palmer, B. A. Holm and R. H. Notter, *Am. J. Physiol.: Lung Cell. Mol. Physiol.*, 2002, **282**, L277-L284.
20. K. W. Lu, J. Perez-Gil and H. W. Taeusch, *Biochim. Biophys. Acta, Biomembr.*, 2009, **1788**, 632-637.
21. C. O. Ciutara and J. A. Zasadzinski, *Soft Matter*, 2021, **17**, 5170-5182.
22. N. J. Balmforth, I. A. Frigaard and G. Ovarlez, *Annu. Rev. Fluid Mech.*, 2014, **46**, 121-146.
23. H. A. Barnes, *J. Non-Newtonian Fluid Mech.*, 1999, **81**, 133-178.

24. P. C. F. Moller, J. Mewis and D. Bonn, *Soft Matter*, 2006, **2**, 274-283.
25. L.-P.-A. Thai, F. Mousseau, E. Oikonomou, M. Radiom and J.-F. Berret, *Colloids Surf. B*, 2019, **178**, 337-345.
26. M. Filoche, C.-F. Tai and J. B. Grotberg, *Proc. Natl. Acad. Sci.*, 2015, **112**, 9287-9292.
27. J. B. Grotberg, M. Filoche, D. F. Willson, K. Raghavendran and R. H. Notter, *Am. J. Respir. Crit. Care Med.*, 2017, **195**, 538-540.
28. E. Herting, C. Härtel and W. Göpel, *Current Opinion in Pediatrics*, 2020, **32**.
29. P. Coussot, L. Tocquer, C. Lanos and G. Ovarlez, *J. Non-Newtonian Fluid Mech.*, 2009, **158**, 85-90.
30. S. Lerouge and J.-F. Berret, in *Polymer Characterization: Rheology, Laser Interferometry, Electrooptics*, eds. K. Dusek and J. F. Joanny, Springer-Verlag, Berlin Heidelberg, 2010, vol. 230, pp. 1-71.
31. M. G. Berni, C. J. Lawrence and D. Machin, *Adv. Colloids Interface Sci.*, 2002, **98**, 217-243.
32. P. Partal, A. J. Kowalski, D. Machin, N. Kiratzis, M. G. Berni and C. J. Lawrence, *Langmuir*, 2001, **17**, 1331-1337.
33. P. Panizza, A. Colin, C. Coulon and D. Roux, *EPJ B*, 1998, **4**, 65-74.
34. L. Chevry, N. K. Sampathkumar, A. Cebers and J. F. Berret, *Phys. Rev. E*, 2013, **88**, 062306.
35. F. Loosli, M. Najm, R. Chan, E. Oikonomou, A. Grados, M. Receveur and J.-F. Berret, *ChemPhysChem*, 2016, **17**, 4134-4143.
36. J.-F. Berret, *Nat. Commun.*, 2016, **7**, 10134.
37. O. El Hamoui, I. Yadav, M. Radiom, F. Wien, J.-F. Berret, J. R. C. van der Maarel and V. Arluison, *Biomacromolecules*, 2020, **21**, 3668-3677.
38. M. Radiom, R. Hénault, S. Mani, A. G. Iankovski, X. Norel and J.-F. Berret, *Soft Matter*, 2021, **17**, 7585-7595.
39. L.-P.-A. Thai, F. Mousseau, E. Oikonomou, M. Radiom and J.-F. Berret, *ACS Nano*, 2020, **14**, 466-475.
40. G. Helgesen, P. Pieranski and A. T. Skjeltorp, *Phys. Rev. Lett.*, 1990, **64**, 1425-1428.
41. H. T. Banks, S. Hu and Z. R. Kenz, *Adv. Appl. Math. Mech.*, 2011, **3**, 1-51.
42. R. G. Larson, *The Structure and Rheology of Complex Fluids*, Oxford University Press, New York, 1998.
43. I. M. Krieger and T. J. Dougherty, *Trans. Soc. Rheol.*, 1959, **3**, 137-152.
44. A. Dörr, A. Sadiki and A. Mehdizadeh, *J. Rheol.*, 2013, **57**, 743-765.
45. K. W. Desmond and E. R. Weeks, *Phys. Rev. E*, 2014, **90**, 022204.
46. S. E. Phan, W. B. Russel, J. X. Zhu and P. M. Chaikin, *J. Chem. Phys.*, 1998, **108**, 9789-9795.